\documentclass[12pt,preprint]{aastex}

\begin{document}
%%%%%%%%%%%%%%%%%%%%%%%%%%%%%%%%%%%%%%%%%%%%%%%%%%%%%%%%%%%%%%%%%%%%%%%%%%%%%%
\newif\ifFigures  %\Figurestrue
\def\Figure#1{\ifFigures{#1}\fi}
\def\figsize{\ifSFB@referee\epsfxsize=0.5\hsize\else\epsfxsize=\hsize\fi}
\def\eq#1{\begin{equation} #1 \end{equation}}
\def\eqarray#1{\begin{eqnarray} #1 \end{eqnarray}}
\def\E#1{\hbox{$10^{#1}$}}
\def\Ca {[25--12]}
\def\Cb {[60--25]}
\def\Cc {[100--60]}
\def\ccd{color--color diagram}
\def\ccds{color--color diagrams}
\def\about  {\hbox{$\sim$}}
\def\la     {\hbox{$\lesssim$}}
\def\ga     {\hbox{$\gtrsim$}}
\def\x      {\hbox{$\times$}}
\def\sub#1{_{\rm #1}}
\def\mic    {\hbox{$\mu$m}}
\def\Mo     {\hbox{$M_{\odot}$}}
\def\Lo     {\hbox{$L_\odot$}}
\def\Mdot{\hbox{$\dot{M}$}}
\def\tV     {\hbox{$\tau_V$}}
\def\half   {\onehalf}
\def\threehalf {\slantfrac{3}{2}}
\def\Ref{\reference{}}
\def\HII{H{\small II}}
\def\n(#1){{}$^{\rm(#1)}$}
\def\comm#1 {{\tt (COMMENT: #1)}}
\def\Lbol {\hbox{$L_{\rm bol}$}}
\def\VT   {\hbox{${\rm V}$}}
\def\BT   {\hbox{${\rm B}$}}
\def\Hip  {\hbox{HIPPARCOS}}
%%%%%%%%%%%%%%%%%%%%%%%%%%%%%%%%%%%%%%%%%%%%%%%%%%%%%%%%%%%%%%%%%%%%%%%%%%%%%%

%\def\Msg{Submitted to ApJL}
%\rightline{\Msg}

\title{Analysis of Stars Common to the IRAS and \Hip\ Surveys}

\author{Timothy G. Knauer\altaffilmark{1}, 
        \v{Z}eljko Ivezi\'c\altaffilmark{2} and
        G.R. Knapp\altaffilmark{2}}

\altaffiltext{1} {Department of Physics and Astronomy, University of Kentucky,
                  Lexington, KY 40506--0055; knauer@pop.uky.edu}
\altaffiltext{2} {Princeton University, Department of Astrophysical Sciences,
                  Princeton, NJ 08544--1001; ivezic,gk@astro.Princeton.edu}

\begin{abstract}
For about 11,000 stars observed in the \Hip\ Survey and detected by IRAS 
we calculate bolometric luminosities by integrating their spectral 
energy distributions from the B band to far-IR wavelengths. 
We present an analysis of the dependence of dust emission on spectral
type and correlations between the luminosity and dust emission for 
about 1000 sources with the best data (parallax error less than 30\%, 
error in luminosity of \about 50\% or better). This subsample includes 
stars of all spectral types and is dominated by K and M giants. 

We use the IRAS [25]-[12] color to select stars with emission from 
circumstellar dust and show that they are found throughout the 
Hertzsprung-Russell diagram, including on the main sequence. Clear 
evidence is found that M giants with dust emission have luminosities 
about 3 times larger (\about 3000 \Lo) than their counterparts without dust, 
and that mass loss on the asymptotic giant branch for both M and C stars 
requires a minimum luminosity of order 2000 \Lo. Above this threshold 
the mass-loss rate seems to be independent of, or only weakly dependent
on, luminosity. We also show that the mass-loss rate for these stars is 
larger than the core mass growth rate, indicating that their evolution is 
dominated by mass loss. 
\end{abstract}


\section{                        Introduction                          }

Dust can be found around pre-main-sequence, main sequence, and 
post-main-sequence stars (e.g. Zuckerman 1980, Habing 1996, Waters \& 
Waelkens 1998). The relationships between the luminosity of a star,
its evolutionary phase, and the properties of its circumstellar dust
are not fully understood. The main obstacle is the lack of a large 
uniform sample which would include both dust emission properties and 
the stellar luminosity and spectral type. In this paper we present such
a sample of stars for which these quantities are obtained by combining 
the results of the IRAS and \Hip\ surveys. While there have been studies 
correlating IRAS data with data from other catalogs (e.g. with the SAO catalog,
Oudmaijer {\em et al.} 1992), the determination of luminosity for a large
number of stars has become possible only recently due to the release of
the \Hip\ astrometric data. 

The Infrared Astronomical Satellite (IRAS) produced a 96\% of the sky
survey  at 12, 25, 60 and 100 \mic,
with the resulting IRAS point source catalog (IRAS PSC) containing over 
250,000 sources. The colors based on IRAS fluxes can efficiently be used to 
distinguish pre-main sequence from post-main sequence stars, and to study 
characteristics of the dust emission (e.g. van der Veen \& Habing 1987, 
Ivezi\'c \& Elitzur 2000, hereafter IE00). The recently released \Hip\ 
(HIgh Precision PARallax COllecting Satellite) catalog contains  
parallaxes of unprecedented accuracy for 118,218 sources, and is 
complete to V\about7.5 (Perryman {\em et al.} 1997). Combining these two data 
sets can yield bolometric luminosities for a large number of stars, and facilitate 
studies of the relationship between the characteristics of dust emission and 
stellar luminosity. 

In Section 2 we describe a catalog of stars obtained by positionally matching 
the IRAS and \Hip\ catalogs. For 11,321 matched sources we calculate bolometric 
fluxes by integrating their spectral energy distributions (SED) from the B band 
to far-IR wavelengths, and with the aid of \Hip\ parallaxes, bolometric luminosities. 
In Section 3, we present an analysis of the dependence of dust emission on spectral
type and correlations between the luminosity and dust emission for 968 sources with 
the most accurate luminosities (standard deviation of \about 50\% or better), and 
in Section 4 we discuss the relationship between mass loss and luminosity for 
asymptotic giant branch (AGB) stars.

\section {   The Synthesis of the IRAS PSC and \Hip\ Catalogs }

We consider only sources from the IRAS PSC with quality flags of at least 2 
in at least 2 IRAS bands\footnote{The reliability of an IRAS flux is described 
by the quality index: 3 -- high, 2 -- low, 1 -- an upper limit.}.  
This is motivated by the results described in IE00 who showed that dusty
stars come in two ``flavors'': AGB stars which typically have fluxes in the 
12 and 25 \mic\ bands larger than the fluxes in the 60 and 100 \mic\ bands,
and various young stellar objects with fluxes in the 60 and 100 \mic\ bands 
larger than the fluxes in the 12 and 25 \mic\ bands. There are 88,619 sources with 
high-quality fluxes in the 12 and 25 \mic\ bands, and 33,435 sources with high quality 
fluxes in the 60 and 100 \mic\ bands. We note that the number of sources with good 
fluxes at only 25 and 60 \mic\ is very small compared to the above (3612, or less
than 3\%), in agreeement with IE00. 

The quoted IRAS positional $2\sigma$ error ellipse for stars is typically 
3\arcsec \x 20\arcsec, 
and its position angle varies across the sky (Beichman {\em et al.} 1988). 
However, the typical distance between \Hip\ sources (about 30\arcmin) is much 
larger than the IRAS error ellipse, and we simply match the positions of stars 
within a circle of radius 25\arcsec. This matching radius ensures that practically 
all true matches are included, while keeping the random association rate below 
$\about 5\times10^{-4}$. Also, this radius is sufficiently small that the number
of Hipparcos sources matched with multiple IRAS sources is negligible (there are
only 4 such cases for which we took the closer match). The positional correlation of 
the subset of $\about$ 125,666 IRAS sources with the full \Hip\ catalog resulted in 
11,321 matches. The positional discrepancies are consistent with the mean IRAS error 
of \about 10\arcsec. From the random association rate we estimate that about 50 matched 
sources (0.4\% of the sample) are random associations\footnote{This number could be 
decreased by about factor 2 by explicitly treating the IRAS error ellipse information, 
but this doesn't seem necessary since the contamination is already negligible.}. 
We estimate the completeness of our matched sample by increasing the matching radius 
to 45\arcsec\ which produces 11,853 matches. Taking into account the increase in the 
expected number of random associations, this implies that our adopted sample includes 
\about 95\% of all true \Hip--IRAS associations.

As expected, the matched sources are predominantly red \Hip\ stars, and the majority 
satisfy $0.9 < \BT-\VT < 1.6$ and $\VT < 9$. When considering the IRAS catalog, 
77\% of sources matched in the \Hip\ catalog are dust-free stars (see Section 3) 
brighter than 1 Jy at 12 \mic. The faint limit for the matched dust-free stars with 
``blue" SEDs is imposed by the IRAS sensitivity. The faint cutoff for matched dusty 
stars (approximately 2 times brighter than the cutoff for the input IRAS sample) with 
``red" SEDs is imposed by the \Hip\  sensitivity. Because of this cutoff, the matching
of the \Hip\ catalog to a deeper IR catalog, e.g. IRAS Faint Source Catalog,
would not significantly increase the sample of dusty stars with good distance 
estimate.

For all matched stars we calculate bolometric fluxes by integrating their SEDs
from the B band to the IRAS 100 \mic\ band. We use photometric measurements provided 
by the \Hip\ main catalog and from the Catalog of Infrared
Observations, which consists of inputs from other catalogs and observations
culled from the literature,  beginning in 1965 (CIO, Gezari {\em et al.} 1993).
The number of flux measurements per star ranges from 6 to \about 20 with a median
value of 10.  We have used 2 integration methods:  linear interpolation in
log($\lambda$) -- log(flux) space and piecewise fitting of the Planck curve. The
two methods typically agree within 20 - 30\%.  From the piecewise fitting of
the Planck curve to the end point pairs we estimate the flux outside the
observed wavelength range to be typically less than 10\%.  From the
bolometric fluxes and \Hip\  parallaxes we calculate bolometric
luminosities for all 11,321 stars.\footnote{We ignore the correction for 
interstellar extinction since the sample is dominated by nearby stars 
($\la$ 300 pc).}

Figure 1 displays the luminosity histogram for a subsample of matched
stars with maximum fractional error in parallax of 30\% and at least 9
fluxes used for the bolometric flux calculation (968 stars). 
Relaxing the parallax cutoff to 50\% roughly doubles the sample size 
but noticeably increases the number of outliers in various diagrams to be 
discussed later. The top
panel in Figure 1 shows the distribution for all stars, irrespective of 
spectral type\footnote{The histogram counts are expressed as normalized
counts per unit log interval (dex$^{-1}$); that is, the integrals of the 
plotted curves over d(logL$_{bol}$) are unity.}.
The error bars are assigned assuming a Poisson error distribution. 
There are two obvious peaks at \about 50 \Lo\ and \about 1000 \Lo. The six panels 
below detail the same information separated by spectral type;  all luminosity classes 
are included.  Since there are so few O and B stars in the sample they 
are treated together. The numbers of stars in each spectral type subsample,
given below the spectral type designation in each panel, show that the 
sample is dominated by K and M giants. The comparison of the histogram for the 
full sample, displayed in the top panel, to the various spectral subsamples, shows that 
the peak at \about 50 \Lo\ is due to K sub-giants, also known as the red clump 
stars (for a detailed discussion see Oudmaijer {\em et al.} 1992), and that the 
peak at \about 1000 \Lo\ is due to M giants (note that the \Hip\ magnitude limit
detects M dwarfs only out to 5 pc).

From the width of the peak for K stars at \about 50 \Lo\ we estimate an upper 
limit on the mean \Lbol\ error of \about 50\% (were this error larger, this 
peak could not be so narrow). Assuming that \Hip\ parallaxes and 
our method for determining bolometric fluxes are not biased with respect to spectral 
type, we adopt this value as the mean \Lbol\ error for the whole sample. However, 
we note that dusty stars with heavily reddened SEDs may have large errors due 
to sparse infrared photometry, and the error distribution in such a case would be
markedly non-Gaussian.

\section{ The Relationship between Luminosity and Spectral Properties}

Figure 2 shows \Lbol\ vs. [25]-[12] color\footnote{We define IRAS [25]-[12] 
color as log(F$_{25}$/F$_{12}$), where F$_{12}$ and F$_{25}$ are the flux
densities taken from the IRAS PSC catalog.} for 913 stars from Figure 1 with IRAS 
flux qualities of at least 2 at 12 and 25 \mic, and with known spectral types. 
Color temperature increases from right 
to left, and [25]-[12] = -0.6 corresponds to the Rayleigh-Jeans tail of the Planck 
function. Stars with this color have no dust, and the distribution width 
indicates the errors in the IRAS fluxes ($\la 10\%$). Stars with [25]-[12] 
$\ga$ -0.5 emit more infrared radiation than a pure black-body; this excess is 
usually attributed to dust emission (however, the infrared excess for some O and B 
stars may be dominated by free-free emission).  A significant fraction (\about 60\%) 
of B and A stars show such infrared excess and are probably young pre-main sequence 
stars (Herbig Ae/Be stars, Waters \& Waelkens 1998). A few F and G stars with
with infrared excess are probably post-asymptotic giant branch stars (Oudmaijer 
{\em et al.} 1992). The M stars are shown in the lower-right panel, and are further
 discussed in \S 4. 
However, it can already be seen in this diagram that M stars with dust are typically
more luminous than those without dust. Note also that early-type stars (OBA) with 
infrared excess have much redder [25]-[12] colors (up to \about 1), than M stars 
with infrared excess ($\la 0$). This difference is due to different dust density 
distributions in the circumstellar envelopes which are flatter for early-type stars
than for late-type stars (IE00).  

Fig. 2 shows that stars with mid-infrared excess can be simply selected as those 
with [25]-[12] $>$ -0.5. Figure 3 displays \Lbol\ vs. \BT-\VT\ diagrams for a subsample
of 947 stars satisfying V$<$11$^{\rm m}$, with (239) and without (708) infrared excess, 
marked by triangles and crosses, respectively. It is easily discernible that dusty 
stars are found throughout the HR diagram. The three dominant dusty populations are 
early type OBA stars with \Lbol $\ga$ 10 \Lo\ and \BT-\VT $\la$ 0.6, and two 
types of late-type stars (further discussed in the following Section) with 
\Lbol \about 10$^3$-10$^4$ \Lo: M stars with \BT-\VT \about 1.3 and carbon stars with 
\BT-\VT $\ga$ 2.0 (Wallerstein \& Knapp 1998).

\section{            Dust around AGB Stars       } 

Asymptotic Giant Branch (AGB) stars are intermediate mass stars
in a late evolutionary stage just preceding the planetary nebula
phase. Due to copious mass loss (up to several 10$^{-4}$ \Mo/yr),
they are surrounded by dusty shells which emit distinctive infrared 
radiation (for an extensive review see Habing 1996). 
It is not clear what the relationship is between the mass-loss rate
and luminosity for AGB stars. Early studies by Gilman (1972) and 
Salpeter (1974), and later by others (e.g. Netzer \& Elitzur, 1993; 
Habing, Tignon, \& Tielens, 1994), showed that the luminosity-to-mass 
ratios for AGB stars are sufficiently large that mass loss could 
be driven by radiation pressure. Furthermore, Ivezi\'c \& Elitzur (1995,
hereafter IE95) 
find that steady-state, radiation pressure driven outflow models can 
explain IRAS colors for at least 95\% AGB stars. However, in these models 
both the mass-loss rate and the stellar luminosity are free parameters, 
and cannot be independently constrained from the observations.
We use the available luminosity and infrared colors for the sample of 
M giants discussed here to study this relationship. 

The top panel in Figure 4 shows the bolometric luminosity vs.
[25]-[12] color diagram for a subsample of 307 M stars with distances
less than 300 pc. The luminosities of 255 stars without dust emission 
([25]-[12] $<$ -0.5) cluster around \Lbol \about 1000 \Lo, while 52 stars 
with dust emission have luminosities on average \about 3 times larger. 
Only 5\% of stars without dust emission have \Lbol $>$ 3000 \Lo, and 
only 13\% of stars with dust emission have \Lbol $<$ 1000 \Lo. This 
difference in bolometric luminosity distributions is better seen in histograms 
shown in the bottom pane, which are plotted separately for each subsample. 
Figure 4 indicates that there is a characteristic luminosity of order 2000 \Lo\ 
for stars to develop a dusty envelope. Above this value the luminosity
does not seem to be correlated with [25]-[12] color and its
median luminosity for dusty M stars is \about 3000 \Lo. 

The scatter of points for dusty M stars around the median luminosity is somewhat larger than 
the expected errors derived in \S3, and thus may be real. However, the procedure 
used to determine bolometric flux is expected to be less accurate for stars 
emitting mostly in the infrared due to sparse flux sampling. The method employed 
here for estimating bolometric flux can be significantly improved by utilizing 
detailed radiative transfer models for stars embedded in dusty envelopes. 
When fitting a model to the observed fluxes, the model SED acts as a smooth
interpolating function which produces a more robust estimate of the
bolometric flux than an arbitrary function such as a piecewise power-law 
(the bolometric flux is simply a scaling parameter, see Ivezi\'c \& Elitzur 1997).
Such a detailed model fitting of a large number of sources will be presented in 
a separate publication.

The difference in median luminosities of stars with and without dust emission
could be due to biased selection procedure. For example, if somehow 
dusty stars with \Lbol $\la$ 1000 \Lo\ are excluded from the sample, then
the median luminosity of the remaining subsample is overestimated. 
We have tested our sample for this and similar possibilities by relaxing 
the constraints on parallax, parallax error, V and F$_{12}$, both individually 
and in various combinations. We find that the difference in median luminosity 
between the two subsamples is a robust result, even when the relaxed selection
cuts result in a three times larger sample. Of course, in this case the scatter 
in luminosity around the median values is also increased, and we find that such 
outliers are mostly stars with relatively large parallax errors. Most notably,
changing the distance cutoff from 200 pc to 500 pc does not effect the median 
luminosities of the two subsamples.

The median luminosity for dusty M stars obtained here (3000 \Lo) is remarkably 
similar to the luminosity of AGB stars detected towards the Galactic bulge 
(\about 2600 \Lo), as determined by Habing {\em et al.} (1985).
This strongly suggests that the luminosity of AGB stars is roughly
the same throughout the Galaxy, and is also the same for stars with different
mass--loss rates (the stars discussed by Habing {\em et al.} are 
significantly redder at IRAS wavelengths than the stars discussed here which were
selected from optical catalogs). It is of interest to 
find out whether this conclusion also holds for AGB stars with carbonaceous 
dust grains.  Due to different optical properties of silicate and carbonaceous
grains, these stars do not show a large increase of [25]-[12] color
for plausible mass-loss rates (IE95), and thus their
mass-loss rate cannot be inferred from their [25]-[12] color. Nevertheless,
the mass-loss rate for C stars can be determined from the intensity of
their CO emission and we utilize such observations to study the relationship
between the mass-loss rate and bolometric luminosity for AGB stars with 
carbonaceous dust grains.
 
Figure 5 shows the mass-loss rate versus bolometric luminosity for a 
sample of 60 C stars with available CO emission observations (Knapp 2000).  
Distances are from the \Hip\ catalog for stars marked as filled circles, and 
indirectly determined (Knapp 2000) for stars marked by open circles.
The mass-loss rate for stars not detected in CO, marked by 
triangles, is determined from their IRAS 60 \mic\ fluxes as described in Jura 
(1991). The distribution of sources shows that for C stars
there seems to be a threshold luminosity of the order 2000 \Lo\ for stars
to develop a dusty envelope. Similarly to the result found for M stars, above 
this threshold the luminosity does not seem to be correlated with 
mass-loss rate. This threshold luminosity of about 2000 \Lo\ is in agreement 
with the theoretical estimate expected for radiatively driven winds 
(e.g. Ivezi\'c, Knapp \& Elitzur 1998, and references therein).

Figure 5 also shows a line corresponding to 
$$\dot M_{core}  = ~ {{L_{bol}}\over{0.007 c^2}},$$
the approximate rate at which the inert helium core is growing from hydrogen
shell burning (Sch\"onberner 1983). At all luminosities 
the mass-loss rate is larger than the core mass growth rate, indicating that 
mass loss dominates the evolution of these stars, in agreement with 
Wallerstein \& Knapp (1998). Assuming the [25]-[12] color 
to mass-loss rate transformation as given by IE95, 
we find that the same conclusion also holds for M stars (c.f. Figure 4).

%---------------------------------------------------------------------------
%                             ACKNOWLEDGEMENTS
\section*{Acknowledgments}
This work has made use of the SIMBAD database and was supported in part 
by NSF grant AST96-18503 to Princeton University. The comments by the 
first anonymous referee, and by Tom Chester, the second referee have 
helped us to considerably improve the paper. We acknowledge Tolya 
Miroshnichenko and Dejan Vinkovi\'c for illuminating discussions. 
%---------------------------------------------------------------------------

%                             REFERENCES

%---------------------------------------------------------------------------

\newpage

%%%%%%%%%%%%%%%%%%%%%%%%%%%%%%%%%%%%%%%%%%%%%%%%%%%%%%%%%%%
\begin{figure}
\vskip -0.5in
\plotone{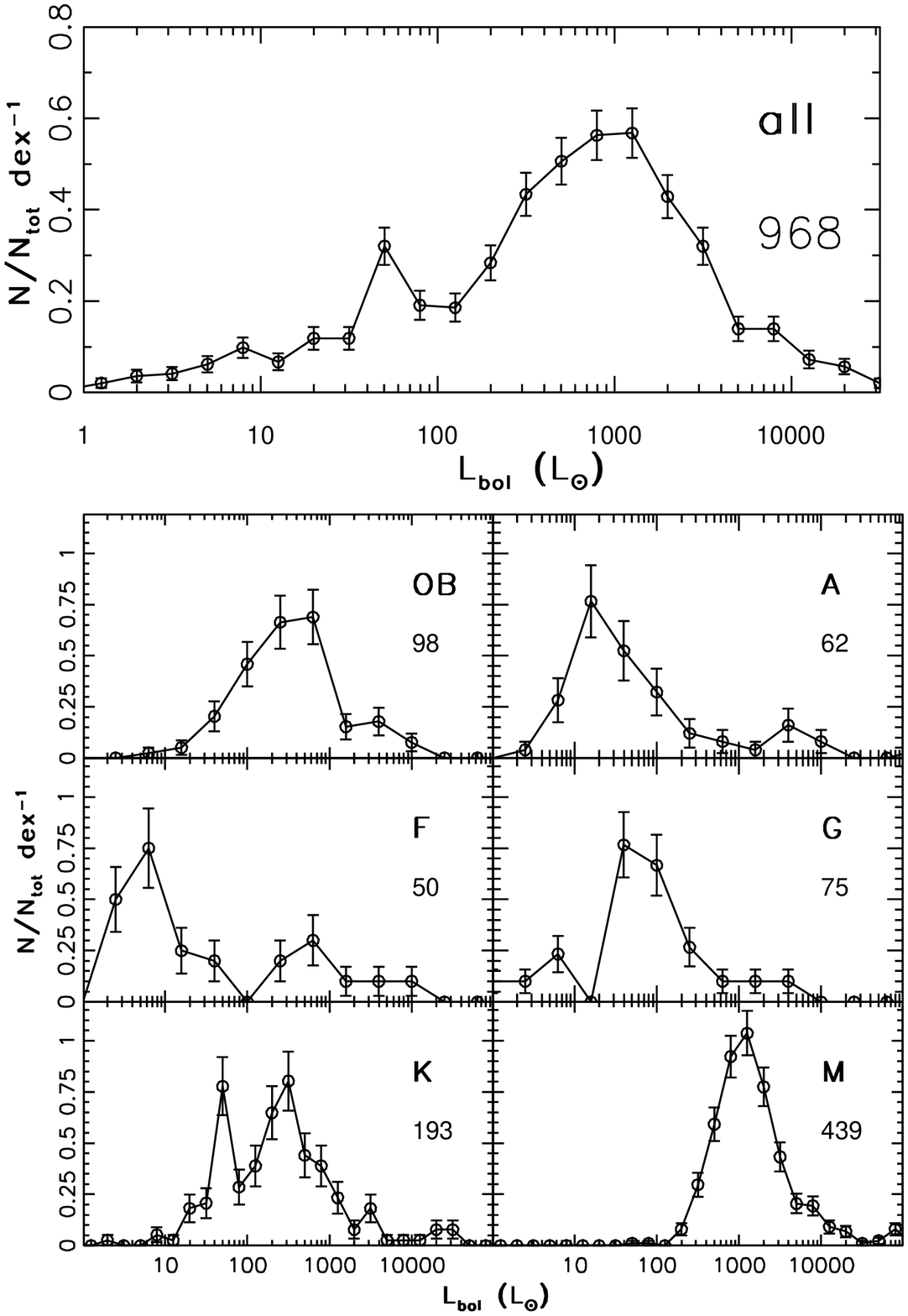}
\vskip -0.4in
\caption{Luminosity histogram for 968 stars with most
reliable luminosity estimates (see text). The top panel shows the 
distribution for all stars irrespective of spectral type,
and the six panels below detail the same information separated by spectral type;  
all luminosity classes are included. The number of stars in each spectral type 
subsample is given below the spectral type designation in each panel.}
\end{figure}
%%%%%%%%%%%%%%%%%%%%%%%%%%%%%%%%%%%%%%%%%%%%%%%%%%%%%%%%%%%% 
\begin{figure}
\vskip -1.5in
\plotone{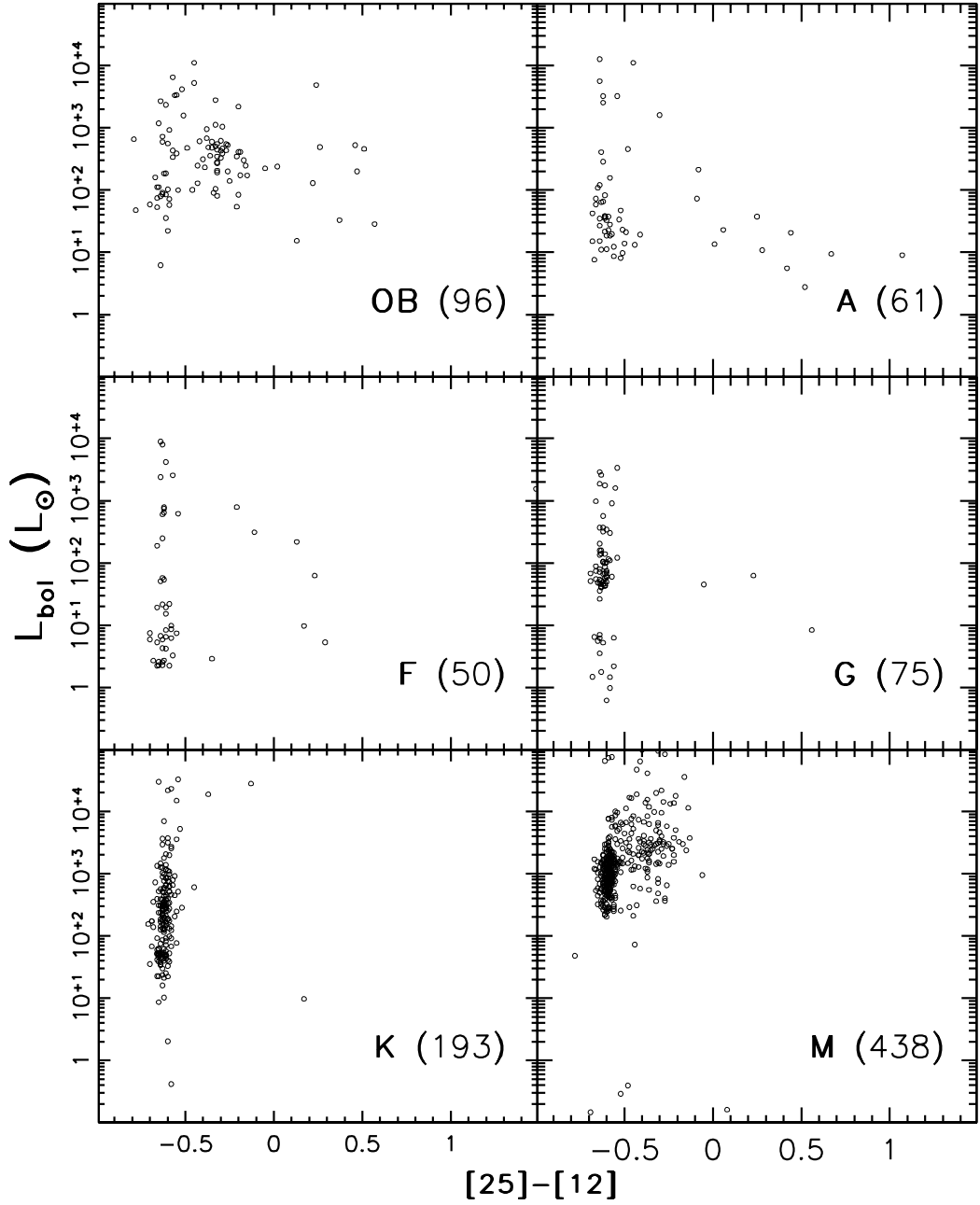}
\caption{Bolometric luminosity -- [25]-[12] diagram for 913 stars,
marked by dots, from Figure 1 which have IRAS flux qualities at 12 and 25 \mic\
of at least 2, and have known spectral types. Panels show different spectral types, 
as marked.}
\end{figure}
%%%%%%%%%%%%%%%%%%%%%%%%%%%%%%%%%%%%%%%%%%%%%%%%%%%%%%%%%%%% 
\begin{figure}
\vskip -2in
\plotone{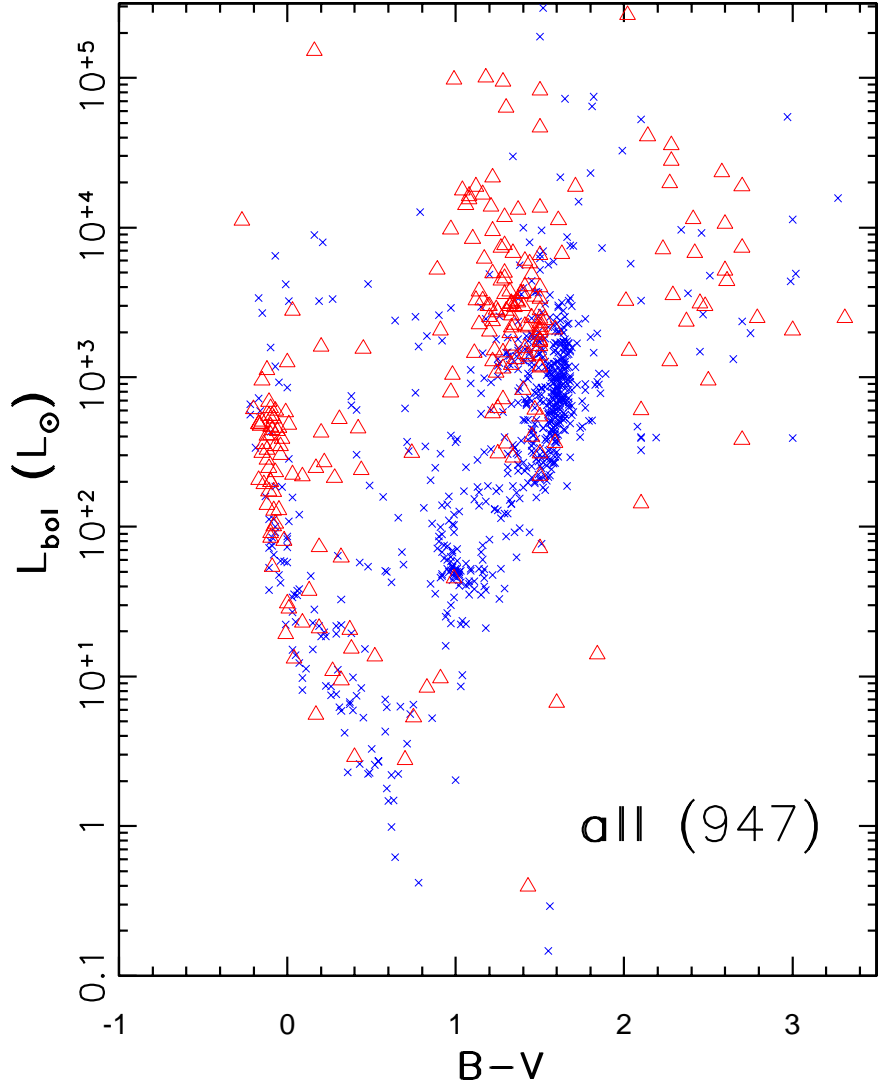}
\caption{Bolometric luminosity -- \BT-\VT\ diagram for 947 stars
from Figure 1 with V$<$11$^{\rm m}$. Triangles: [25]-[12] $>$ -0.5. Crosses: 
[25]-[12] $<$ -0.5}
\end{figure}
%%%%%%%%%%%%%%%%%%%%%%%%%%%%%%%%%%%%%%%%%%%%%%%%%%%%%%%%%%%% 
\begin{figure}
\vskip -0.2in
\plotone{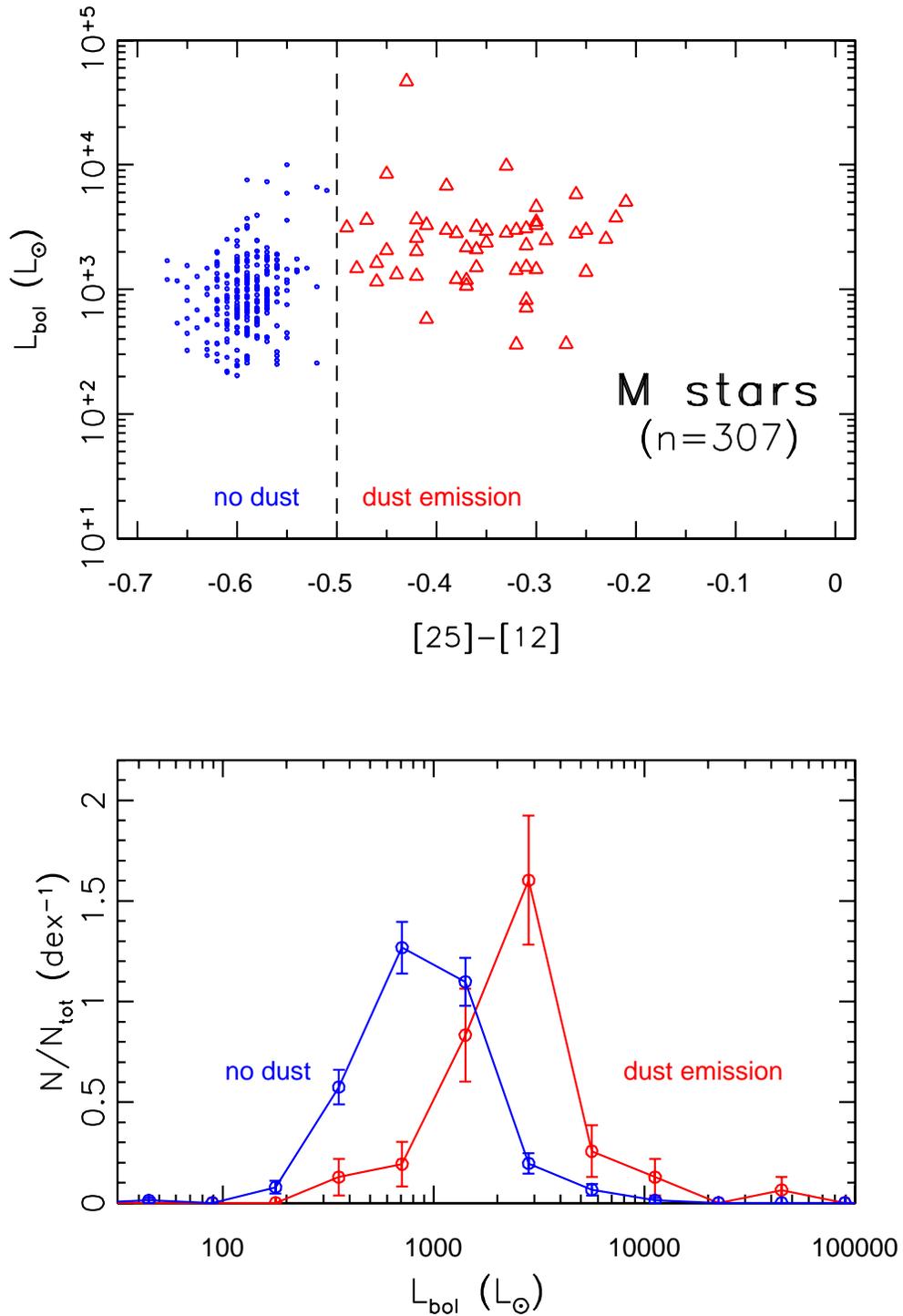}
\vskip -0.6in
\caption{Top panel: bolometric luminosity -- [25]-[12] diagram for 
307 M giants with distances less than 300 pc. Bottom panel: luminosity histogram 
for non-dusty vs. dusty M giants, as implied by their [25]-[12] color.}
\end{figure}
%%%%%%%%%%%%%%%%%%%%%%%%%%%%%%%%%%%%%%%%%%%%%%%%%%%%%%%%%%%% 
\begin{figure}
\vskip -2in
\plotone{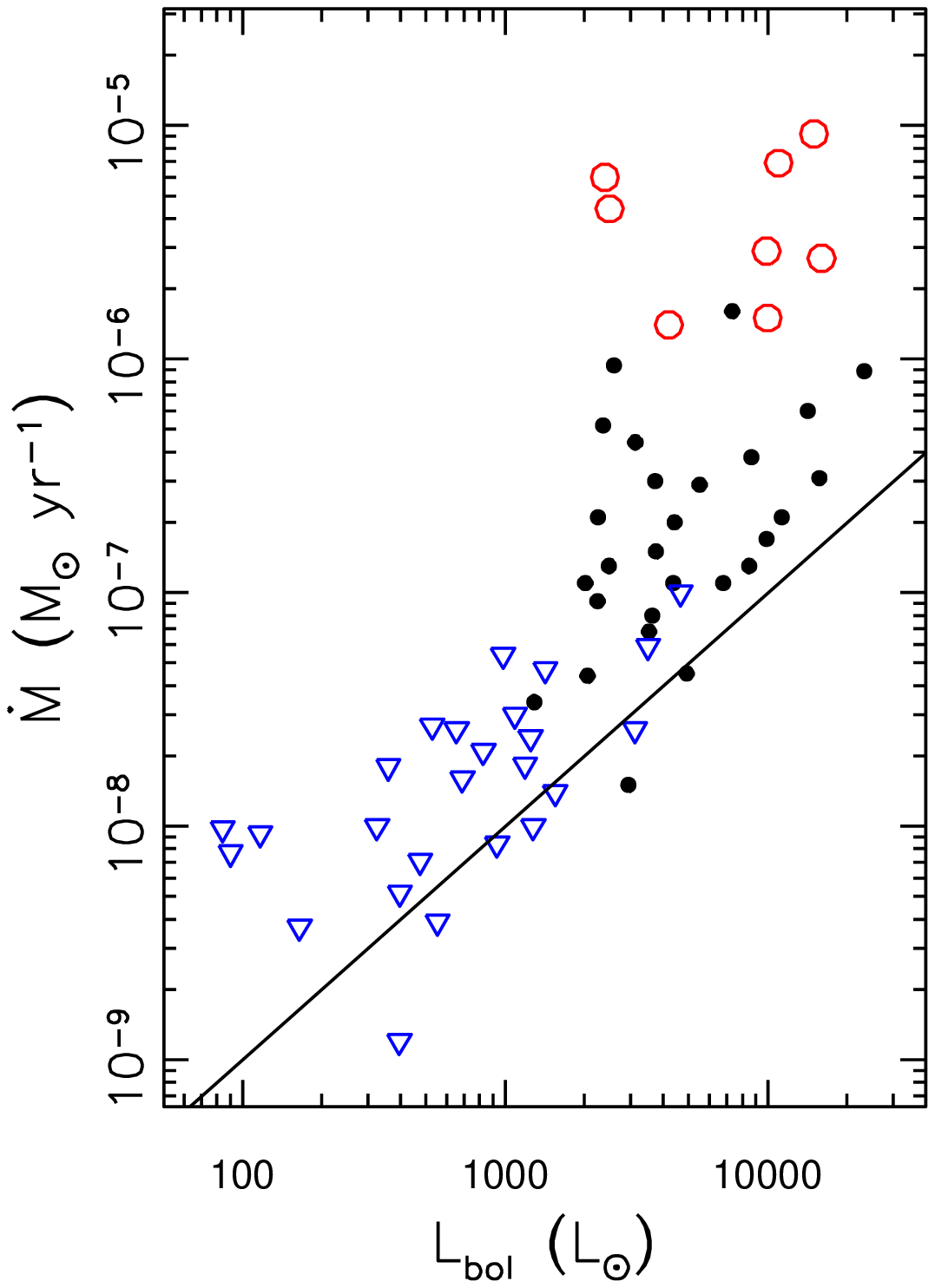}
\caption{Mass-loss rate vs. bolometric luminosity for C stars (various  
symbols correspond to different methods for estimating mass-loss rate and distance, 
see text). The line indicates the core mass growth rate.}
\end{figure}
%%%%%%%%%%%%%%%%%%%%%%%%%%%%%%%%%%%%%%%%%%%%%%%%%%%%%%%%%%%% 


\begin{references}   
\Ref Beichman, C.A., Neugebauer, G., Habing, H.J., Clegg, P.E., \& Chester, 
     T.J. 1988, {\em Infrared astronomical satellite (IRAS) catalogs and atlases. 
     Volume 1: Explanatory supplement}, (Washington, DC: US GPO)
\Ref Gezari, D.Y., Schmitz, M., Pitts, P.S., \& Mead, J.M. 1993, Catalog of 
     Infrared Observations (NASA RP-1294), 3rd ed. (Washington: NASA)
\Ref Gilman, R.C. 1972, ApJ, 178, 423
\Ref Habing, H. 1996, A\&A Rev., 7, 97
\Ref Habing, H., {\em et al.} 1985, A\&A, 152, L1 
\Ref Habing, H.J., Tignon, J. \& Tielens, A.G.G.M. 1994, A\&A 286, 523.
\Ref Ivezi\'c, \v Z., \& Elitzur, M. 1995, ApJ, 445, 415 (IE95)
\Ref Ivezi\'c, \v Z., \& Elitzur, M. 1997, MNRAS, 287, 799
\Ref Ivezi\'c, \v Z., \& Elitzur, M. 2000, ApJ, 534, L93 (IE00)
\Ref Ivezi\'c, \v Z., Knapp, G.R., \& Elitzur M. 1998, Proceedings of 
     the 6$^{th}$ Annual Conference of the CFD Society of Canada, June 1998, 
     Qu\'ebec, p. IV-13; also astro-ph/9805003
\Ref Jura, M. 1991, A\&A Rev. 2, 227.
\Ref Knapp, G.R. 2000, submitted to ApJ.
\Ref Netzer N., \& Elitzur M. 1993, ApJ, 410, 701.
\Ref Oudmaijer, R.D., {\em et al.} 1992, A\&AS, 96, 625.
\Ref Perryman, M.A.C., {\em et al.} 1997, A\&A Letters, 323, L49.
\Ref Salpeter, E.E. 1974, ApJ, 193, 579 
\Ref Sch\"onberner, D. 1983, ApJ 272, 708
\Ref van der Veen W.E.C.J., \& Habing H.J. 1988, A\&A, 194, 125
\Ref Wallerstein, G., \& Knapp, G.R. 1998, ARA\&A, 36, 369
\Ref Waters, L.B.F.M., \& Waelkens, C. 1998, ARA\&A, 36, 233
\Ref Zuckerman, B. 1980, ARA\&A, 18, 263
\end{references}
\end{document}